\documentclass[11pt]{article}
\usepackage{setspace}
\usepackage{amsfonts}
\usepackage{amssymb}         
\usepackage{amsmath}    
\usepackage{multirow}
\usepackage{graphicx}
\usepackage{algorithm}
\usepackage{algorithmic}

\newtheorem{theorem}{Theorem}
\newtheorem{lemma}{Lemma}

\newtheorem{proposition}{Proposition}
\newtheorem{remark}{Remark}

\newtheorem{definition}{Definition}

\newcommand{\Eb}{{\mathbf{E}}}

\newcommand{\Ncal}{\mathcal{N}}

\begin{document}

\title{Mean-Variance Hedging for Pricing European Options Under Assumption of 
Non-continuous Trading}
\author{\textsc{Vladimir Nikulin\thanks{Email: vnikulin.uq@gmail.com}}    \\ 
Department of Mathematics, University of Queensland \\
Brisbane, Australia
}
\date{}
\maketitle
\thispagestyle{empty}

\begin{abstract}
We consider a portfolio with call option and the corresponding
underlying asset under the standard assumption that stock-market price
represents a random variable with lognormal distribution.
Minimizing the variance (\textit{hedging risk}) of the portfolio on
the date of maturity of the call option we find a fraction of the
asset per unit call option.  As a direct consequence we derive
the statistically fair lookback call option price in explicit form.

In contrast to the famous Black-Scholes theory, any 
portfolio can not be regarded as risk-free because no additional
transactions are supposed to be conducted over the life of the
contract, but the sequence of independent portfolios will reduce
risk to zero asymptotically. This property is illustrated in the
experimental section using a dataset of daily stock prices of 18
leading Australian companies for the period of 3 years.
\end{abstract}

\section{Introduction} \label{sec:intro}

A typical asset $S$ as a geometric Brownian motion process
\cite{Sam65}, \cite{Sam73}, \cite{Sam73a} has its price governed by the following
equation:
$$\frac{dS}{S} = \mu dt + \sigma dz,$$
where $\mu \in R$ and $\sigma \in R_+$ are \textit{appreciation}
and \textit{volatility} coefficients, $z$ is a standard Wiener
process with $$\Eb dz =0, \Eb (dz)^2 =dt.$$

According to the Ito's lemma,
\begin{equation} \label{eq:ito}
   d\log{\{S(t)\}} =\left( \mu - \frac{\sigma^2}{2}\right) dt +\sigma dz.
\end{equation}

Therefore,
$$\log{\{S(t+T)\}} \sim \Ncal(\log{\{S(t)\}}+
\left( \mu - \frac{\sigma^2}{2}\right)T, \sigma\sqrt{T}),$$ where
$\Ncal(a, b)$ is the distribution function of a normal random variable
with mean $a$ and standard deviation $b.$ Let us denote the
corresponding density by
$$f_S(x) = \frac{1}{ \sqrt{2\pi} \cdot x \cdot b(T)}
\exp{\{ - \frac{(\log{\{x\}} - a(T))^2}{2b^2(T)} \}},$$ where
$$a(T)
= \log{\{S(t)\}} + (\mu-\frac{\sigma^2}{2})T,
  b(T)=  \sigma \sqrt{T}.$$

\begin{definition} \label{df:def}
A European call option contract allows its owner to purchase one
unit of the underlying asset at a fixed price $K$  after date  $t
+ T$  in the future or the owner of the call option may decide not
to exercise if the price of the underlying asset is less than
strike price $K$. Respectively, the value of a European call option 
with maturity date $t + T$ is
$$C(t+T) = \psi(S(t+T)-K) = \max{\{0, S(t+T)-K\}}.$$
\end{definition}

The fundamental problem in mathematical finance \cite{HanJac98}
is to find the fair hedger or price of such an option at a time
$t$ prior to expiry.

\subsection{Expectations hedging and Black-Scholes formula}

According to \cite{Mer73} and \cite{Whit01} one suggestion would
be that
\begin{equation} \label{eq:rule}
   C_{exp}(t) = e^{-rT} \cdot \Eb \psi(S(t+T)-K)
\end{equation}
where $r$ is the riskless rate.

\begin{proposition} \label{pr:bsprop}
Suppose that parameters $t, T, r$ and $K$ are arbitrary fixed.
Then,
\begin{equation} \label{pr:expect}
 e^{-rT} \cdot \Eb \psi(S(t+T)-K) = S(t) \cdot e^{(\mu-r)T} \cdot
\Phi(\alpha) - K \cdot e^{-rT} \cdot \Phi(\beta)
\end{equation}
where $\Phi$ is a distribution function of the standard normal
law, and
$$\alpha = \frac{\log{\{\frac{S(t)}{K}\}}+(\mu+\frac{\sigma^2}{2})T}{\sigma\sqrt{T}},
\beta = \alpha - \sigma \sqrt{T}.$$
\end{proposition}

The formula (\ref{pr:expect}) was proved by \cite{Sam73} who also
noted that (\ref{pr:expect}) will coincide with Black-Scholes
formula \cite{BlSch73} in the particular case $\mu = r:$
\begin{equation} \label{pr:BS}
 C_{BS}(t) = S(t) \cdot \Phi(\alpha_r) - K \cdot e^{-rT} \cdot \Phi(\beta_r),
\end{equation}
where
$$\alpha_r = \frac{\log{\{\frac{S(t)}{K}\}}+(r+\frac{\sigma^2}{2})T}{\sigma\sqrt{T}},
\beta_r = \alpha_r - \sigma \sqrt{T}.$$

\begin{remark} Similar results for variance gamma processes may be
found in \cite{MadCar98}. Also, we note papers \cite{CarMad99}
and \cite{BorNovikov} where formulas for call options were
obtained using methods based on Fourier transformations.
\end{remark}

However, suggestion (\ref{eq:rule}) ignores the fact that the
seller can himself continue to trade actively on the stock market.
The $BS$-formula (\ref{pr:BS}) provides a unique price of the
European contingent claim \cite{KarKou96} in an ideal, complete
and unconstrained market. Under these conditions the contract is
self-financing and risk-free, to seller as well to buyer. In the
given financial market, the mean-variance hedging problem in
continuous time \cite{Duffie91}, \cite{Sch92} is to find for a given
payoff a best approximation by means of a self-financing trading
strategies where the optimality criterion is the expected squared
error \cite{BobSch04}. In a series of recent papers, this problem
has been formulated and treated as a linear-quadratic stochastic
control problem, see for instance 
\cite{Hend05}, \cite{Bia02}, \cite{BiaGua02}.

\begin{figure}[t]
\begin{center}
\includegraphics[scale=0.75]{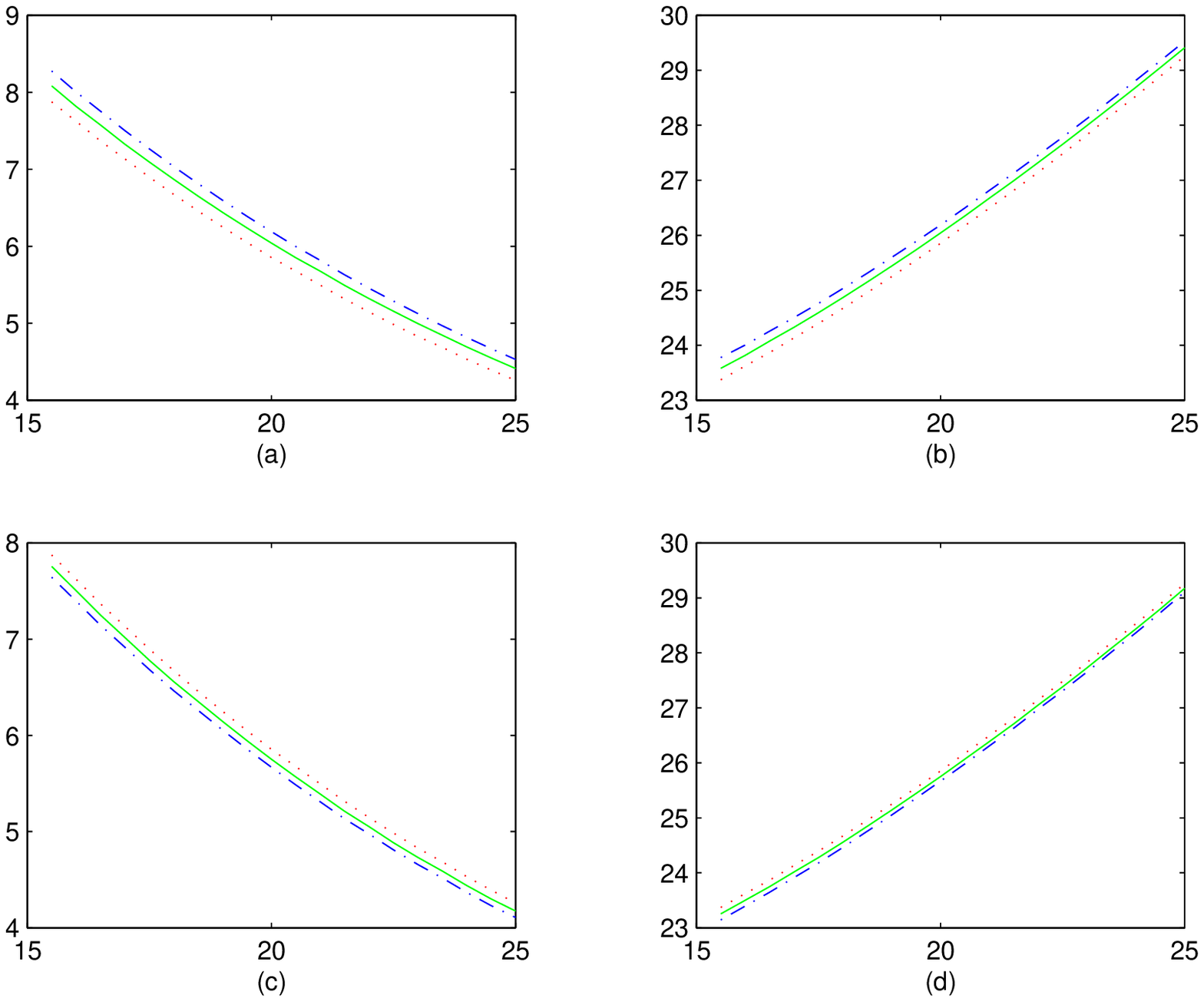}
\end{center}
  \caption{The left column illustrates behavior of call option $C$,
the right column illustrates behavior of
  the sum of call option and strike price $K$ as a function of $K$. The following parameters
were used:
  (a-b) $S(t) =20, \mu = 0.1, r =0.05, \sigma =1, T=180/365;$
  (c-d) $S(t) =20, \mu = 0.02, r =0.05, \sigma =1, T=180/365;$
   green solid line, blue dash-dot line and red dotted line correspond to $MV$
  (\ref{eq:slt1}), Expectations (\ref{pr:expect}) and $BS$ (\ref{pr:BS})
solutions, respectively.} \label{fig: figure1}
\end{figure}

According to \cite{Whit01} the original Black-Scholes formula is
criticized on the grounds that it holds out the quite unrealistic
prospect of risk-free operation, that it can sacrifice asset
maximization to exact meeting of the contract.

\section{Mean-variance hedger}

Let us consider portfolio $F$ consisting of the option $C$ and $h$
units of the underlying asset $S$. The value of the portfolio
(seller case) is therefore
\begin{equation} \label{eq:portf}
  F(t) = -C(t) +h \cdot S(t)
\end{equation}
or
\begin{displaymath}  \label{eq:repres}
   F(t) =  \left\{
  \begin{aligned}
     (h - 1) \cdot S(t) + K
     \hspace{0.08in} \mbox{if} \hspace{0.05in} S(t) \geq K; \\
      h \cdot S(t),
     \hspace{0.08in} \mbox{otherwise}.
  \end{aligned} \right.
\end{displaymath} 

\begin{figure}[t]
\begin{center}
\includegraphics[scale=0.75]{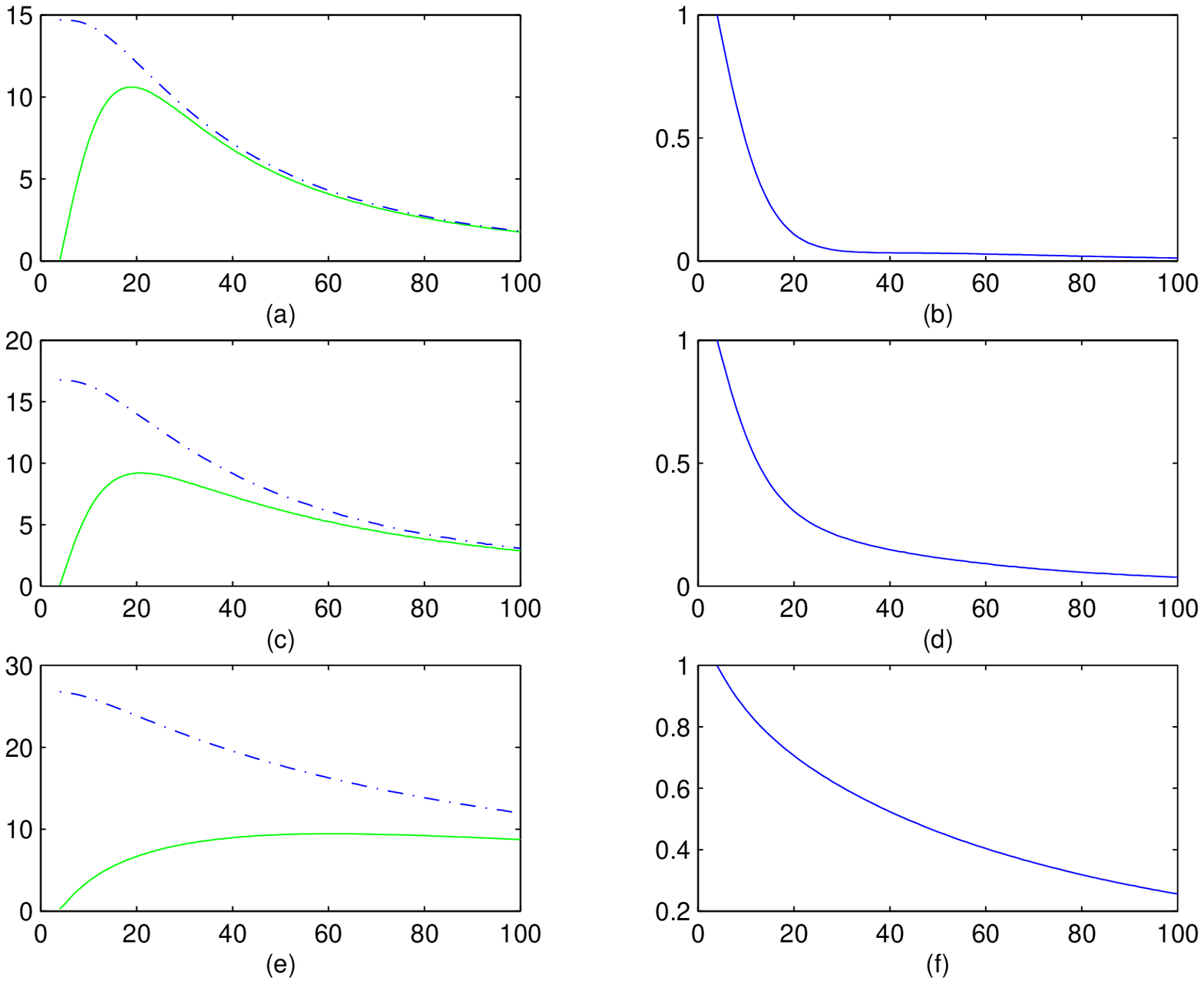}
\end{center}
  \caption{The left column:
  standard deviation of the portfolio as a function of $K$
  where the green solid line corresponds to (\ref{eq:slt1}),
  and the blue dash-dot line corresponds to the portfolio with riskless asset only
  (\ref{eq:rule}) (Expectations approach);
  the right column: value of the parameter $h$ as a function of strike price $K$.
  The following parameters were used: a-b) $\mu=0.1, \sigma=0.9$; c-d) $\mu=0.1, \sigma=1.0$;
  e-f) $\mu=0.1, \sigma=1.4$. All other parameters are the same as in the case of
  Figure~\ref{fig: figure1}.}
\label{fig: figure2}
\end{figure}

According to the fundamental principles of mean-variance model
\cite{Mark52} we consider the rule that the investor  
considers expected return as a desirable thing and variance of
return as an undesirable thing. The above rule may be implemented
using different methods. For example, we can define fractions of
the portfolio by maximizing the ratio of expected return to the
standard deviation of the portfolio, or we can minimize variance
assuming that the expected return is fixed. In our case we will
minimize the variance of the portfolio (\ref{eq:portf}) assuming
that the number of call options is arbitrary fixed. In order to
simplify notations we consider a portfolio with one call option.

The following Theorem represents the main result of the paper. It
will establish the value of the parameter $h$ in order to minimize
variance of the portfolio (\ref{eq:repres}). Then, we find hedging
 call option price or simply \textit{hedger}
\begin{equation} \label{eq:border}
    C_{MV}(t) =  h \cdot S(t) - e^{-rT} \Eb F(t+T).
\end{equation}

\begin{theorem} \label{th:main}
Suppose that the portfolio $F$ is defined in (\ref{eq:repres}).
Then, the hedging problem
$$ \min_h Q_{var}(F(t+T)) \hspace{0.05in} \mbox{where}
\hspace{0.05in} Q_{var}(F(t+T)) := \Eb \left[ F(t+T) - \Eb
F(t+T)\right]^2$$ has the unique solution
\begin{equation} \label{eq:slt}
  h = \frac{A_4 - K \cdot A_2 + (A_2+A_3)(K \cdot A_1 - A_2)}{A_4+A_5-(A_2+A_3)^2},
\end{equation}
where
$$A_1(K) := \int_K^{\infty} f_S(x) dx = \Phi(\frac{a(T) - \log K }{b(T)});$$
$$A_2(K) := \int_K^{\infty} x f_S(x) dx =
\exp{\{a(T)+\frac{b^2(T)}{2}\}}\Phi(b(T)+\frac{a(T) - \log
K}{b(T)});$$
$$A_3(K) := \int_0^{K} x f_S(x) dx$$
$$= \exp{\{a(T)+\frac{b^2(T)}{2}\}}
    \left(1 - \Phi(b(T) + \frac{a(T) - \log K}{b(T)})\right);$$
$$A_4(K) := \int_K^{\infty} x^2 f_S(x) dx =
\exp{\{2(a(T)+b^2(T))\}}\Phi(2 b(T)+\frac{a(T) - \log K}{b(T)});$$
$$A_5(K) := \int_0^{K} x^2 f_S(x) dx$$
$$ = \exp{\{2(a(T)+b^2(T))\}}
  \left(1- \Phi(2 b(T)+\frac{a(T) + \log K}{b(T)})\right).$$
\end{theorem}
{\it Proof:} According to the definition of variance (\textit{hedging
risk})
\begin{equation} \label{eq:varn}
  Q_{var}(F(t+T)) = \Eb F^2(t+T) - \left( \Eb F(t+T) \right)^2,
\end{equation}
where
$$\Eb F^2(t+T) = (h-1)^2 A_4 +2K(h-1)A_2+K^2A_1+h^2A_5;$$
$$\Eb F(t+T) = h(A_2+A_3)-A_2+K \cdot A_1.$$

Minimizing (\ref{eq:varn}) as a function of $h$ we find the required
solution (\ref{eq:slt}). $\blacksquare$

\begin{figure}[t]
\includegraphics[scale=0.75]{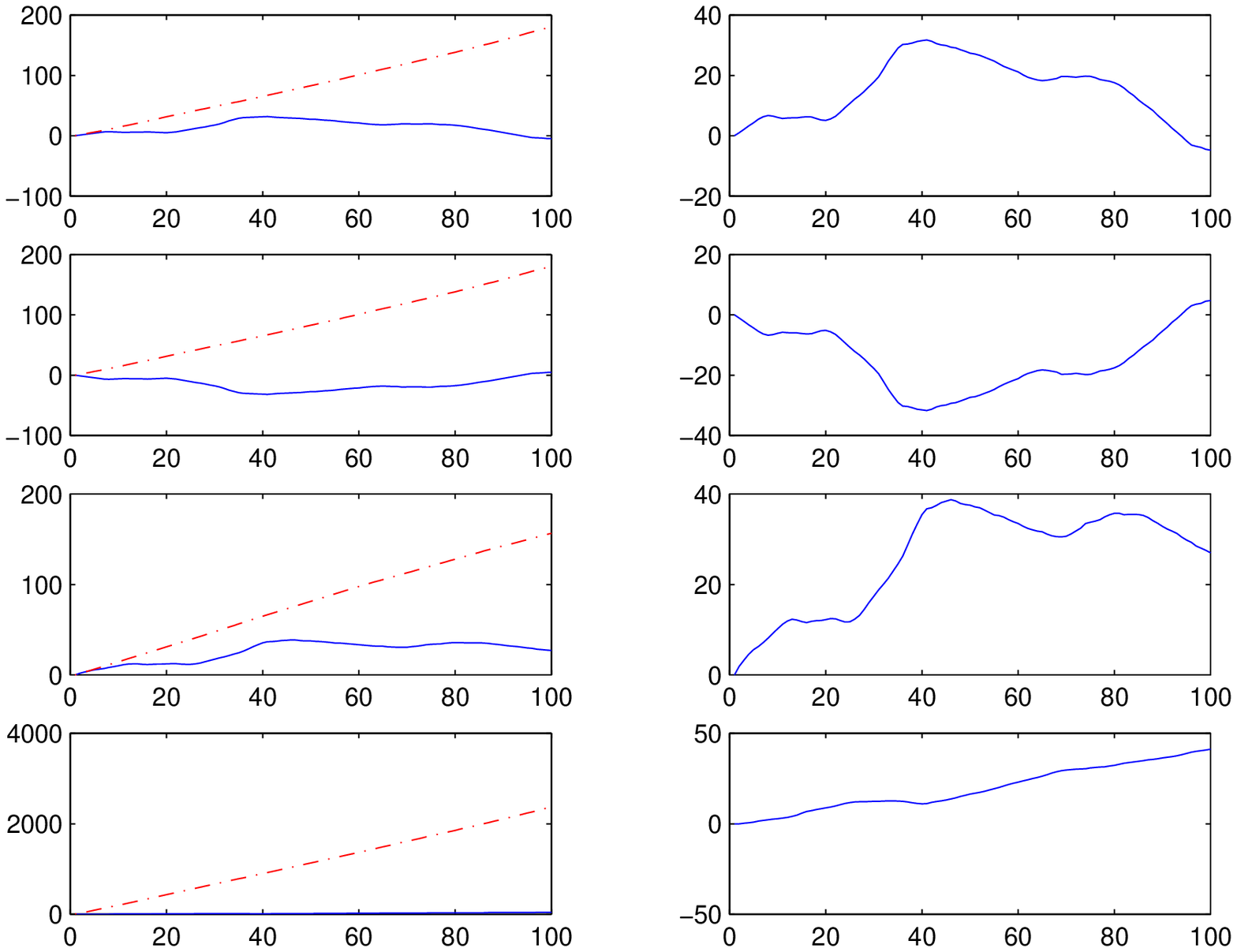}
  \caption{General case of $m = 18$ assets (see Table~\ref{tb:table1}).
  Performance of portfolio based on Expectations (1st and 2nd rows) and
  $MV$ (3rd and 4th rows) Algorithms where solid line represents profits of buyer
(1st and 3rd rows) and seller
  (2nd and 4th rows), dash-dotted line represents an average turnover.
  All computations were done according to (\ref{eq:average} - \ref{eq:taverage}) and
(\ref{eq:aver}).} \label{fig: figure5}
\end{figure}

Finally, we find a mean-variance hedger according to
(\ref{eq:border})
\begin{equation} \label{eq:slt1}
  C_{MV}(t) = h \cdot S(t) - e^{-rT} \left[ h(A_2+A_3) -A_2 + K \cdot A_1 \right]
\end{equation}
where parameter $h$ is defined in (\ref{eq:slt}).

Next, we can re-write (\ref{pr:expect}) using new notations which
were introduced in this section:
\begin{equation} \label{eq:slt1a}
  C_{exp}(t) = e^{-rT} \left(A_2 -K \cdot A_1 \right).
\end{equation}

The above call option relates to the portfolio combined with riskless
asset only:
\begin{equation} \label{eq:portf1}
  F(t) = -C(t) +h,
\end{equation}
where $h$ is a constant parameter. The corresponding standard
deviation is invariant under $h$ and is given by the following
formula (see Figure~\ref{fig: figure2}):
$$S_{dev}(F(t+T)) = \sqrt{A_1(1-A_1)K^2+2A_2K(A_1-1)+A_4-A^2_2}.$$

\begin{remark} Note that the price (\ref{eq:slt1}) may be negative, in contrast to
the price (\ref{eq:slt1a}) which is always positive by definition.
\end{remark}

Using relations $$A_2 +A_3 = \exp{\{a+0.5 b^2\}}, A_4 +A_5 =
\exp{\{2(a+b^2)\}},$$ we can simplify (\ref{eq:slt}):
\begin{equation} \label{eq:mvs}
  h(K) = \frac{A_4 - K \cdot A_2 + \exp{\{a+0.5 b^2\}}(K \cdot A_1 - A_2)}
  {\exp{\{2a + b^2\}} (\exp{\{b^2\}}-1)}.
\end{equation}

Let us consider some marginal properties of the coefficients
$A_i(K), i=1..5:$
$$A_1(K) \underset{K \rightarrow 0}{\longrightarrow} 1,
  A_3(K) \underset{K \rightarrow 0}{\longrightarrow} 0,
  A_5(K) \underset{K \rightarrow 0}{\longrightarrow} 0.$$

It follows from above that
$$h(K)  \underset{K \rightarrow 0}{\longrightarrow} 1
  \hspace{0.08in} \mbox{and} \hspace{0.08in}
  S_{dev}(K)  \underset{K \rightarrow 0}{\longrightarrow} 0
 \hspace{0.05in} \mbox{(see Figure~\ref{fig: figure2}).}$$

\begin{proposition} \label{pr:hprop}
Assuming that $\sigma > 0,$ the following range $0 < h < 1$ is
valid where asset fraction parameter $h$ is defined in
(\ref{eq:slt}).
\end{proposition}

The proof of the above Proposition~\ref{pr:hprop} follows from the following two 
Lemmas.
\begin{lemma} Assuming that $\Phi$ is a distribution function of standard normal law
the following relation is valid: 
\begin{equation} \label{eq:lem1}
  \frac{\Phi(v+b) - \Phi(v)}{\Phi(v) - \Phi(v-b)} < \exp{\{0.5 b^2 - bv\}}, 
\end{equation}
for any  $v \in R$ and  $b \in R_{+}.$
\end{lemma}
{\it Proof:} We have
$$\Phi(v) - \Phi(v-b) = \frac{1}{\sqrt{2\pi}} \int_v^{v+b} \exp{\{-\frac{(t-b)^2}{2}\}} dt =
\frac{e^{-0.5 b^2}}{\sqrt{2\pi}} \int_v^{v+b} e^{-0.5 t^2}
e^{bt}dt$$
$$< \frac{e^{-0.5 b^2 + bv}}{\sqrt{2\pi}} \int_v^{v+b} e^{-0.5 t^2} dt =
  \frac{e^{-0.5 b^2 + bv}}{\sqrt{2\pi}} \left[\Phi(v+b) - \Phi(v) \right].$$
Therefore, the proof is completed. $\blacksquare$

\begin{lemma} Assuming that $\Phi$ is a distribution function of standard normal law
the following relation is valid:
\begin{equation} \label{eq:lem2}
  \Phi(v) < \frac{e^{b^2}\Phi(v+b) + 
  \exp{\{0.5 b^2 - bv\}}\Phi(v-b)}{1+\exp{\{0.5 b^2-bv\}}}, 
\end{equation}
for any  $v \in R$ and  $b \in R_{+}.$
\end{lemma}
{\it Proof:} We have
$$e^{b^2} \Phi(v+b) - \Phi(v) > e^{b^2} \left[ \Phi(v+b) - \Phi(v) \right] =
\frac{e^{b^2}}{\sqrt{2\pi}} \int_v^{v+b} \exp{\{-\frac{t^2}{2}\}}
dt$$
$$= \frac{e^{b^2}}{\sqrt{2\pi}} \int_{v-b}^{v} e^{-0.5 (t+b)^2} dt =
  \frac{e^{0.5b^2}}{\sqrt{2\pi}} \int_{v-b}^{v} e^{-0.5 t^2 - b t}
  dt$$
$$  > \exp{\{0.5b^2-b v\}} \left[\Phi(v) - \Phi(v-b) \right].$$
Therefore, the proof is completed. $\blacksquare$

\begin{figure}[t]
\begin{center}
\includegraphics[scale=0.75]{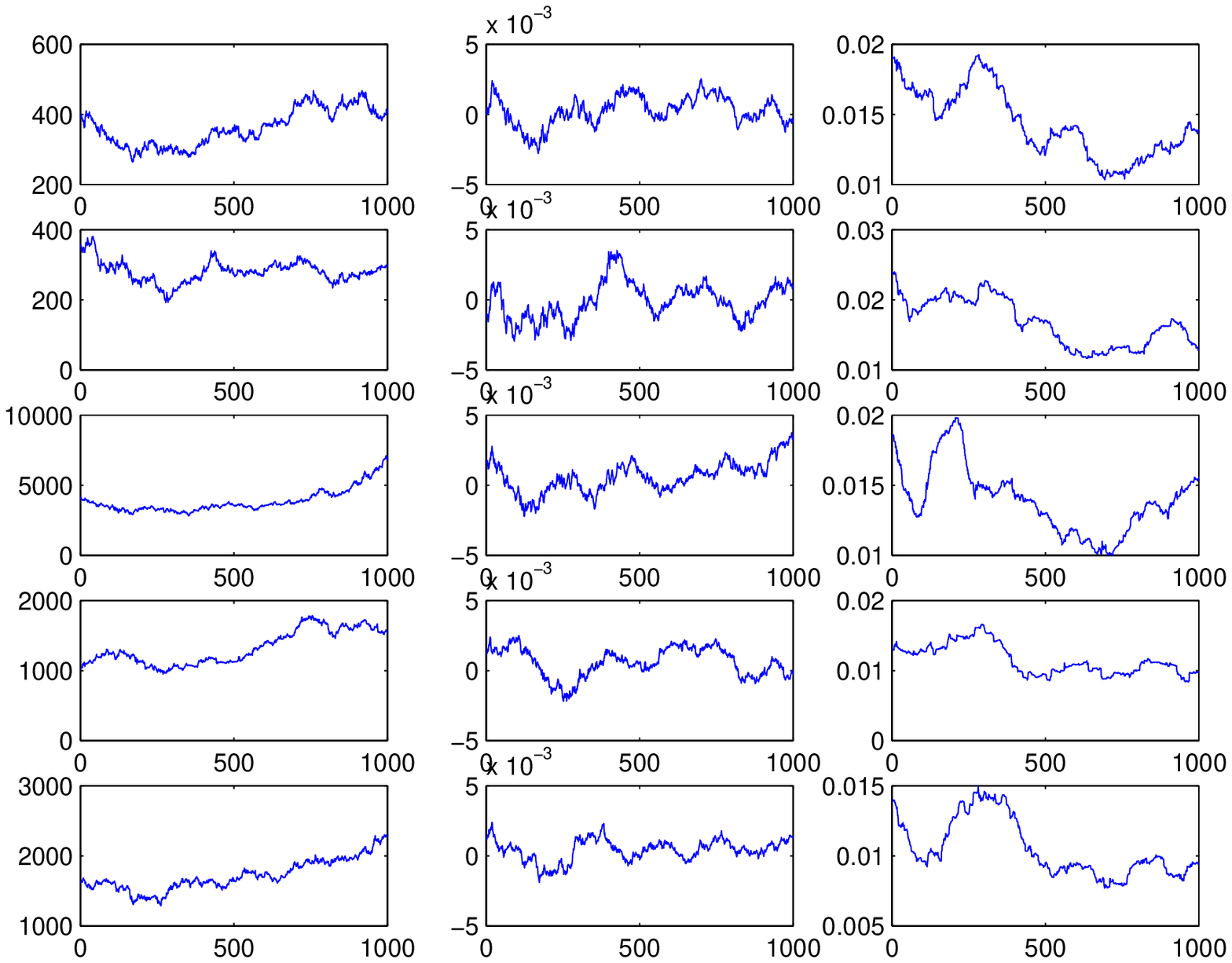}
\end{center}
  \caption{The left column illustrates stock market prices of Fairfax, Harvey Norman,
  Rio-Tinto, Tabcorp and Westpac for the period of 1000 days ending on 10th January 2006;
  the middle and right columns illustrate moving means and standard deviations which 
  were computed
  according to (\ref{eq:mu}) and (\ref{eq:sigma}) using smoothing parameter $n  =120$.}
\label{fig: figure3}
\end{figure}

Using definition of the coefficients $A_i(K), i=1..5,$ we can
re-write (\ref{eq:mvs}) in the following form: 
$$h = \frac{e^{b^2}\Phi(v+b) -\Phi(v) - \exp{\{0.5 b^2- b v \}}
  \left[ \Phi(v) - \Phi(v-b)\right]}{e^{b^2}-1},$$
where $v=b + \frac{a-\log{\{K\}}}{b}.$

Then, a strict upper and a lower bounds for $h$ (as it is stated
in the Proposition~\ref{pr:hprop}) follows from (\ref{eq:lem1})
and (\ref{eq:lem2}) if $\sigma >0.$

\begin{remark} The left column of the Figure~\ref{fig: figure1} illustrates
descending property of the hedging call option price as a function
of the strike price. It is interesting to note that the sum of the
call option price and strike price is an ascending function of the
strike price. This fact is quite explainable because the second
part of the transaction (purchase of the stock) is not compulsory.
According to the Figure~\ref{fig: figure1} the formulas
(\ref{pr:expect}) and (\ref{eq:slt1}) are more flexible comparing
with Black-Scholes formula which is independent of the appreciation
coefficient $\mu.$
\end{remark}

\section{Experiments}

Based on the representation (\ref{eq:ito}) we can formulate
an estimator for the historical (\textit{moving}) volatility
\cite{McM93}:
\begin{equation} \label{eq:sigma}
  \hat{\sigma}_{i,t} = \sqrt{\frac{\sum_{j=1}^n \left( R_{i,t-j} - 
\overline{R}_{i,t} \right)^2}{n-1}}
\end{equation}
where $$R_{i,t-j} =\log{\frac{S_{i, t-j+1}}{S_{i, t-j}}};$$
$S_{i,t}$ is a closing price of $i$-asset on the day $t > n$ and
$$\overline{R}_{i,t} = \frac{1}{n} \sum_{j=1}^n 
\log{\frac{S_{i,t-j+1}}{S_{i,t-j}}}.$$ 
Then, we can estimate historical (\textit{moving}) appreciation:
\begin{equation} \label{eq:mu}
  \hat{\mu}_{i,t} = \overline{R}_{i,t} + \frac{1}{2} \hat{\sigma}^2_{i,t}.
\end{equation}

\subsection{Expectations hedging}

\begin{figure}[t]
\includegraphics[scale=0.75]{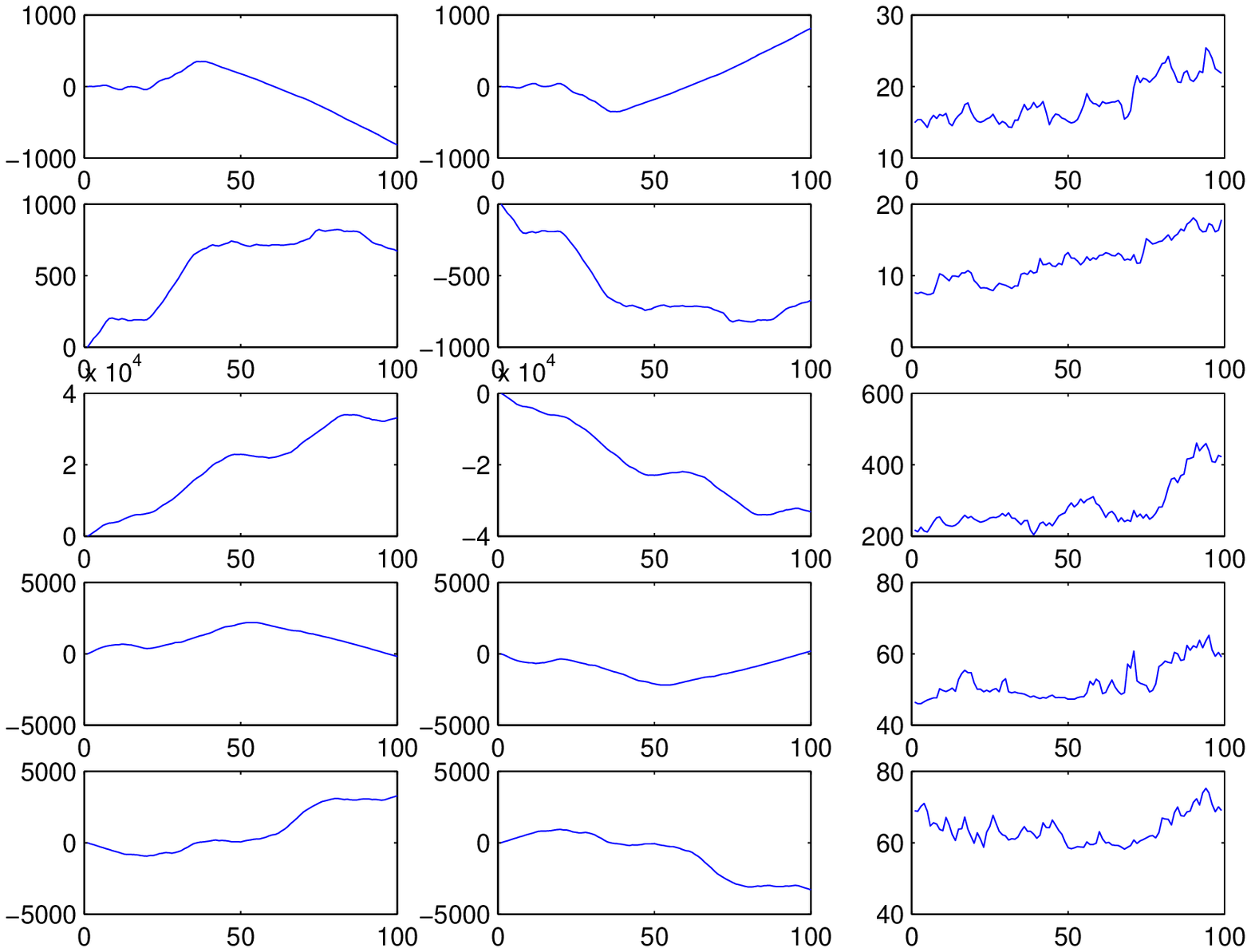}
  \caption{Expectations approach: the first 2 columns represent profits of
buyer (\ref{eq:profb}) and seller
  (\ref{eq:profs}) during period of 100 days ending on 10th January 2006;
  the third column represents corresponding call options
  price (\ref{pr:expect}).}
\end{figure}   \label{fig: figure4}

The call option price $C_{i,t}$ (see the right column of the
Figure~\ref{fig: figure4}) was computed according to
(\ref{eq:slt1a}) subject to the following condition: $C_{i,t} \geq
0.03 \cdot S_{i,t} \hspace{0.05in}
 \mbox{(administrative fees of not less than 3 \%).}$
The strike price was computed using historical appreciation and
volatility coefficients
$$K_{i,t} = S_{i,t} \cdot \exp{\{\frac{\mu_{i,t} \cdot T}{\beta +
\gamma \cdot \sigma_{i,t}} \}}, \beta = 1.1, \gamma = 20.$$

\begin{table}[t]  \singlespacing
\caption{All prices are given in cents. The first column gives the name of
asset, the second column gives the average 
  price during period of 100 days ending on 10th January 2006.
  Columns 3-6 represent final profits of buyer and seller in cases of
 Expectations and $MV$ approaches.}
\begin{center}  \scriptsize
 \begin{tabular}{llllll}  
   \hline \noalign{\smallskip}
\multicolumn{1}{l}{Asset} \hspace{0.2in} &
\multicolumn{1}{l}{Average} \hspace{0.2in} &
\multicolumn{2}{l}{Expectations} &
\multicolumn{2}{l}{Mean-Variance}\\
   \hline  \noalign{\smallskip}
Name  \hspace{0.2in} & Price \hspace{0.2in} & Buy \hspace{0.1in} & Sell \hspace{0.2in} & Buy \hspace{0.1in} & Sell \\
 \noalign{\smallskip}
\hline
 \noalign{\smallskip}
ANZ Bank    &   2306.53 &   922.63  &   -922.63 &   1801.59 &   3125.66 \\
CBA,    Commonwealth Bank   &   3909.16 &   -1410.77    &   1410.77 &   -485.93 &   5702.04 \\
CML,    Coles Myers &   990.6   &   24.78   &   -24.78  &   461.42  &   1477.59 \\
DJS,    David Jones &   236.11  &   1428.42 &   -1428.42    &   1844.33 &   -513.65 \\
FXJ,    Fairfax &   417.18  &   -816.73 &   816.73  &   -657.82 &   -58.73  \\
HVN,    Harvey Norman   &   277.67  &   671.77  &   -671.77 &   347.12  &   642.38  \\
NAB,    National Bank   &   3193.27 &   -4433.27    &   4433.27 &   -4172.65    &   6730.61 \\
PBL,    Publish.Brodcast.   &   1625.36 &   3416    &   -3416   &   2419.06 &   2119.44 \\
QAN,    Qantas  &   350.14  &   2234.44 &   -2234.44    &   1344.69 &   64.72   \\
QBE Insurance   &   1813.72 &   1216.89 &   -1216.89    &   5023.97 &   1466.49 \\
RIO,    Rio-Tinto   &   5788.09 &   33156.66    &   -33156.66   &   47506.18    &   -10089.97   \\
STO,    Santos  &   1141.14 &   -9401.15    &   9401.15 &   -2086.32    &   3910.58 \\
TAH,    Tabcorp &   1606.01 &   -183.37 &   183.37  &   -1558.01    &   2335.71 \\
TEN Network &   342.78  &   -1192.97    &   1192.97 &   -1816.85    &   818.41  \\
TLS,    Telstra &   412.26  &   -1372.16    &   1372.16 &   -1924.31    &   -1068.74    \\
WBC,    Westpac Bank    &   2104.37 &   3293.2  &   -3293.2 &   2949.07 &   2272.26 \\
WOW,    Woolworth   &   1633.53 &   -3596.63    &   3596.63 &   -2517.34    &   3440.02 \\
WPL,    Woodside Petroleum  &   3361.07 &   -21427.93   &   21427.93    &   3208.89 &   11465.86    \\
\hline
 \end{tabular}
\end{center}
\end{table}  \label{tb:table1}

The left and middle columns of Figure~\ref{fig: figure4} correspond to
the profit of buyer $PB_{i,t}$ and seller $PS_{i,t}$ which were
computed for the 100 consecutive days ending on 10th January 2006
$(j=0..100)$. The computations were conducted using the following
rules:
$$PB_{i,t+T+j+1} =  PB_{i,t+T+j}$$
\begin{displaymath} \label{eq:profb}
 + \left\{
  \begin{aligned}
     S_{i,t+T+j+1} - K_{i,t+j+1} -C_{i,t+j+1}
     \hspace{0.08in} \mbox{if} \hspace{0.05in} S_{i,t+T+j+1} \geq K_{i,t+j+1}; \\
     -C_{i,t+j+1}, \hspace{0.08in} \mbox{otherwise};
  \end{aligned}  \right.
\end{displaymath}
and
$$PS_{i,t+T+j+1} =  PS_{i,t+T+j}$$
\begin{displaymath} \label{eq:profs}
   + \left\{y
  \begin{aligned}
     K_{i,t+j+1} +C_{i,t+j+1} - S_{i,t+T+j+1}
     \hspace{0.08in} \mbox{if} \hspace{0.05in} S_{i,t+T+j+1} \geq K_{i,t+j+1}; \\
     C_{i,t+j+1}, \hspace{0.08in} \mbox{otherwise}, 
  \end{aligned}  \right.
\end{displaymath}
where initial values of  $PB_{i,t+T}$ and $PS_{i,t+T}$ are set to
zero.

In order to estimate the performance of the system against the whole
set of $m$ assets, we computed average stock-prices $u_i, i=1..m,$
for the period under consideration. Then, we computed weights $w_i
\propto \left( u_i \right)^{-1}, \sum_{i=1}^m w_i =1.$

The average profits of buyers $AB_{t}$ and sellers $AS_{t}$ were
computed using the following formulas: 
\begin{subequations}
\begin{align}
  \label{eq:average}
  AB_{t} = \sum_{i=1}^m w_i \cdot PB_{i,t}, \\
  \label{eq:saverage}
  AS_{t} = \sum_{i=1}^m w_i \cdot PS_{i,t}, \\
  \label{eq:taverage}
  AT_{t} = \sum_{i=1}^m w_i \cdot C_{i,t}
\end{align}
\end{subequations}
where dash-dotted line corresponds to the average turnover
$AT_{t}$ (see the first two lines of the Figure~\ref{fig: figure5}).

\subsection{Mean-variance hedging}

\begin{figure}[t]
\includegraphics[scale=0.75]{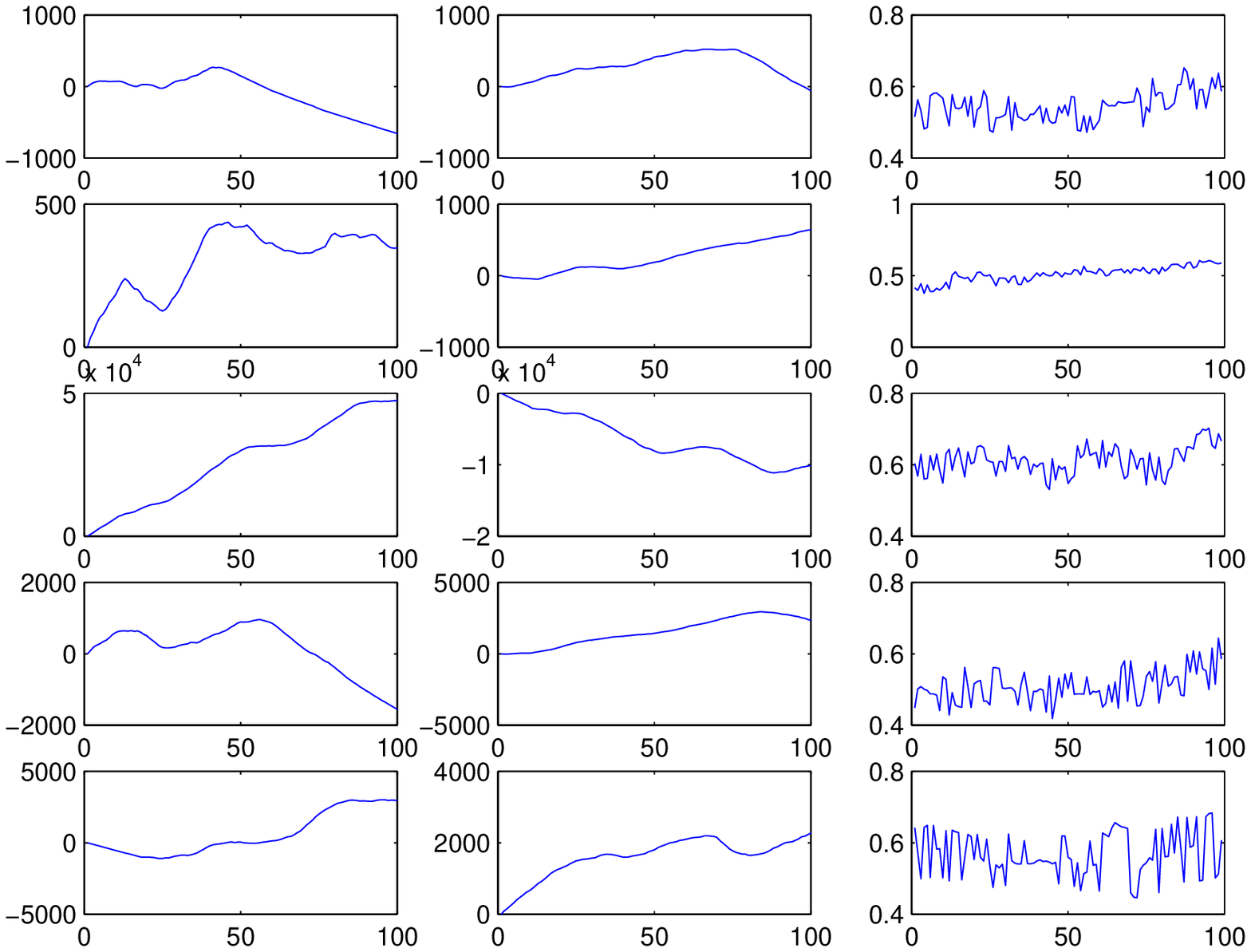}
  \caption{$MV$ approach: the first 2 columns represent profits of buyer
(\ref{eq:profb}) and seller
  (\ref{eq:profits}) during period of 100 days ending on 10th January 2006;
the third column represents $h$ parameter (\ref{eq:slt}).}
\end{figure}   \label{fig:figure6}

Here we make modifications of (\ref{eq:profs}) and
(\ref{eq:taverage}) (all other formulas remain the same as in the
previous Section):
$$  PS_{i,t+T+j+1} =  PS_{i,t+T+j} + C_{i,t+j+1}$$
\begin{displaymath} \label{eq:profits}
   + \left\{
  \begin{aligned}
       (1-h_{i,j+1})(K_{i,t+j+1}-S_{i,t+T+j+1})
     \hspace{0.08in} \mbox{if} \hspace{0.05in} S_{i,t+T+j+1} \geq K_{i,t+j+1}; \\
       h_{i,j+1} (S_{i,t+T+j+1}-S_{i,t+j+1}), \hspace{0.08in} \mbox{otherwise},
\end{aligned} \right.
\end{displaymath}   
and
\begin{equation}  \label{eq:aver}
     AQ_{t} = \sum_{i=1}^m w_i \cdot (C_{i,t}+h_{i,t} S_{i,t}).
\end{equation}
As a result, turnover of seller will be larger comparing with
turnover of buyer. Third and Fourth lines of the Figure~\ref{fig:
figure5} illustrate average profits of buyer and seller.

\section{Concluding remarks}

The classical equation (\ref{eq:rule}) establishes the hedging so
that a transaction will be statistically profitable for buyer if
price is smaller, or profitable for seller if price is higher. Any
particular transaction is not risk-free, but the sequence of
independent transactions may reduce risk essentially (see for
details Figure~\ref{fig: figure5} and Table~1.

In contrast, risk-free formula (\ref{pr:BS}) was obtained under
ideal assumption of absolute liquidity of the market. It means,
any transaction represents a continuous sequence of trading, which
(as it was noticed in many papers) can-not be achieved in real
terms.

Combination of the call option with the corresponding asset represents
an additional degree of flexibility. On the one hand, it will help
to reduce risk for seller. On the other hand, the call option price
will be reduced in the case if performance of the stock is good
historically. Anyway, in accordance with $MV$ approach a seller
will calculate hedging call option price somewhere between prices
computed according to the $BS$ and Expectations approaches.
Therefore, $MV$ hedger may be regarded as a compromise between 2
base solutions (see Figure~\ref{fig: figure1}).

Comparing the third and fourth lines with first two lines of the
Figure~\ref{fig: figure5}, which were developed using the same
regulation parameters, we can see advantages of the $MV$ approach
against Expectations approach.


\begin{thebibliography}{19}

\bibitem{Bia02}
F. Biagini.
\lq\lq Mean-Variance Hedging for Interest Rate Models with Stochastic Volatility."
{\em Decisions in Economics and Finance}, vol. 25, pp. 1-17, 2002.

\bibitem{BiaGua02}
F. Biagini and P. Guasoni.
\lq\lq Mean-Variance Hedging with Random Volatility Jumps."
{\em Stochastic Analysis and Applications}, vol. 20(3), pp. 471-494, 2002.

\bibitem{BlSch73}
F. Black and M. Scholes.
\lq\lq The pricing of options and corporate liabilities."
{\em Journal of Political Economy}, vol. 81, pp. 637-659, 1973.

\bibitem{BobSch04}
O. Bobrovnytska and M. Schweizer.
\lq\lq Mean-variance hedging and stochastic control: beyond the Brownian setting."
{\em IEEE Transactions on Automatic Control}, vol. 49(3), pp. 396-408, 2004.

\bibitem{BorNovikov}
K. Borovkov and A. Novikov.
\lq\lq On a new approach to calculating expectations for option pricing."
{\em Journal of Applied Probability}, vol. 39, pp. 889-895, 2002.

\bibitem{CarMad99}
P. Carr and D. Madan.
\lq\lq Option valuation using the fast Fourier transform."
{\em Journal of Computational Finance}, vol. 2, pp. 61-73, 1999.

\bibitem{Duffie91}
D. Duffie and H. Richardson.
\lq\lq Mean-Variance Hedging in Continuous Time."
{\em The Annals of Applied Probability}, vol. 1(1), pp. 1-15, 1991.

\bibitem{HanJac98}
D. Hand and S. Jacka.
\lq\lq Statistics in Finance." {\em Arnold}, 1998.

\bibitem{Hend05}
V. Henderson.
\lq\lq Analytical Comparisons of Option Prices in Stochastic Volatility Models."
{\em Mathematical Finance}, vol. 15(1), pp. 49-59, 2005.

\bibitem{KarKou96}
I. Karatzas and S. Kou.
\lq\lq On the pricing of contingent claims under constraints." 
{\em Journal of Applied Probability}, vol. 6(2), pp. 321-369, 1996.

\bibitem{MadCar98}
D. Madan, P. Carr and E. Chang.
\lq\lq The variance gamma process and option pricing."
{\em European Finance Review}, vol. 2, pp. 79-105, 1998.

\bibitem{Mark52}
H. Markowitz.
\lq\lq Portfolio selection."
{\em The Journal of Finance}, vol. 7(1), pp. 77-91, 1952.

\bibitem{McM93}
L. McMillan.
\lq\lq Options as a strategic investment: a comprehensive analysis 
of listed option strategies." 
{\em New York Institute of Finance}, 1993.

\bibitem{Mer73}
R. Merton.
\lq\lq Theory of rational option pricing."
{\em The Bell Journal of Economics and Management Science}, vol. 4(2), pp. 141-183, 1973.

\bibitem{Sam65}
P. Samuelson.
\lq\lq Rational theory of warrant pricing."
{\em Industrial Management Review}, vol. 6(2), pp. 13-31, 1965.

\bibitem{Sam73}
P. Samuelson.
\lq\lq Mathematics of speculative price."
{\em SIAM Review}, vol. 15(1), pp. 369-374, 1973.

\bibitem{Sam73a}
P. Samuelson.
\lq\lq Proof that properly discounted present values of assets vibrate randomly."
{\em The Bell Journal of Economics and Management Science}, vol. 4(2), pp. 369-374, 1973.

\bibitem{Sch92}
M. Schweizer.
\lq\lq Mean-Variance Hedging for General Claims."
{\em The Annals of Applied Probability}, vol. 2(1), pp. 171-179, 1992.

\bibitem{Whit01}
P. Whittle.
\lq\lq On the structure of proper Black-Scholes formulae."
{\em Journal of Applied Probability}, vol. 38A, pp. 243-248, 2001.

\end{thebibliography}
\end{document}